\documentclass[aps,prl,twocolumn,superscriptaddress]{revtex4-2}
\usepackage{amsmath}
\usepackage{graphicx}

\newcommand{\fzpc}{f_{\rm ZPC}}
\newcommand{\fthe}{f_{\rm Theory}}
\newcommand{\feq}{f_{\rm eq}}
\newcommand{\pT}{p_{\rm T}}
\newcommand{\lsub}{l_{\rm sub}}
\newcommand{\mpT}{\langle\pT\rangle}
\newcommand{\vpT}{{\rm var}\left(\pT\right)}
\newcommand{\MD}{{\rm RMD}}
\newcommand{\tca}{t_{c,1}}
\newcommand{\tcb}{t_{c,2}}
\newcommand{\appropto}{\mathrel{\vcenter{
  \offinterlineskip\halign{\hfil$##$\cr
    \propto\cr\noalign{\kern2pt}\sim\cr\noalign{\kern-2pt}}}}}

\begin{document}

\title{Effectiveness of parton cascade in
  solving the relativistic Boltzmann equation in a box}

\author{Todd Mendenhall}
\affiliation{Department of Physics, East Carolina University,
  Greenville, NC 27858, USA}
\affiliation{Applied Research Associates, Inc., Raleigh, NC 27615, USA}
\author{Zi-Wei Lin}
\email{linz@ecu.edu}
\affiliation{Department of Physics, East Carolina University, Greenville, NC 27858, USA}

\date{\today}

\begin{abstract}
We benchmark the ZPC parton cascade with an exact analytical solution of
the relativistic Boltzmann equation for a homogeneous and massless gas
with a constant and isotropic elastic cross section.
We measure the accuracy of ZPC with the relative mean deviation
between its momentum distribution and the exact solution. 
We use two generalized collision schemes to further improve
the accuracy of ZPC over the recent $t$-minimum collision scheme.
We find that ZPC can reproduce very well the time evolution of
the single-particle distribution function for the exact solution's
initial condition, with one generalized collision scheme giving an
accuracy better than $1\%$ for the momentum distribution at any time
in all studied cases, including very high opacities where naively the
parton cascade approach is expected to fail.
\end{abstract}

\maketitle

\section{Introduction}

A new phase of matter with parton degrees of freedom called the
quark-gluon plasma (QGP) has been formed in high-energy nuclear
collisions at the Relativistic Heavy Ion Collider~\cite{STAR:2005gfr}
and the Large Hadron Collider~\cite{ALICE:2013nwm}. 
The STAR Collaboration has shown that the matter formed
in Au+Au collisions at $\sqrt{s_{\rm NN}}=3$ GeV collisions is
dominated by hadronic interactions~\cite{STAR:2021yiu}, while 
by $\sqrt{s_{\rm NN}}=4.5$ GeV there is evidence~\cite{STAR:2025owm}
for the onset of dominant partonic interactions from the observation
of the constituent quark number scaling~\cite{Molnar:2003ff}. 
There has also been evidence that the QGP is even formed in small
collision systems such as d+Au collisions at
RHIC~\cite{PHENIX:2014fnc,PHENIX:2018lia,STAR:2022pfn} as well as p+p
and p+Pb collisions at the LHC~\cite{CMS:2010ifv,CMS:2015yux}. 
It is thus important to study small systems or systems at lower
energies to establish the origin of collectivity 
and to reliably extract the QGP properties. 
The hydrodynamics approach has been very successful in 
describing large systems at high energies; however, it is not clear
whether it can be applied to small or low energy systems
~\cite{Heller:2015dha,Kurkela:2018ygx,Kurkela:2019set}. 
For a small system, we expect the non-equilibrium
dynamics to be important, such as the parton escape
mechanism~\cite{He:2015hfa,Lin:2015ucn,Li:2024ivj} for the generation
of anisotropic
flows. Transport models~\cite{Zhang:1997ej,Xu:2004mz,Kurkela:2022qhn} 
and effective kinetic theory~\cite{Kurkela:2018vqr} 
are natural tools to address non-equilibrium dynamics. 

Therefore,  it is important to be able to accurate model 
the parton dynamics of the QGP. 
At very high temperatures, quarks and gluons are weakly
interacting due to asymptotic freedom, where the QGP could be
pictured as a relativistic gas as described by the relativistic
Boltzmann equation (RBE). The RBE gives the evolution of the
single-particle distribution function $f\left(x^\mu,p^\mu,t\right)$
and has a wide range of applications from heavy ion
collisions~\cite{Sorge:1989dy,Zhang:1997ej,Bass:1998ca,Xu:2004mz,Kurkela:2022qhn,Kurkela:2018vqr}
to core-collapse supernovae~\cite{Janka:2006fh}. 
Because the RBE is a complicated non-linear integro-differential
equation, exact analytical solutions of the RBE are rare. 
Numerical or approximate solutions of the RBE are more common,
such as solutions under the relaxation time 
approximation~\cite{Florkowski:2013lza}.

Zhang's parton cascade (ZPC)~\cite{Zhang:1997ej} is a Monte Carlo
program that  uses the cascade method to numerically 
solves the RBE under two-to-two parton scatterings. 
In general, parton cascades suffer from causality violation at high
densities~\cite{Kodama:1983yk,Kortemeyer:1995di,Tornkvist:2023kan},
which arises due to the geometric interpretation of the cross section
$\sigma$. Thus, one naively expects ZPC to be accurate only in the
dilute limit when the interaction length $\sqrt{\sigma/\pi}$ is much
smaller than the mean free path $l$, i.e., when $\chi \ll 1$ with
$\chi=\sqrt{\sigma/\pi}/l$ being the opacity~\cite{Zhang:1998tj}. 
The parton subdivision method can be used to reduce the inaccuracies
that originate from the causality violation~\cite{Molnar:2000jh}. 
Parton subdivision can be understood in terms of the transformation 
$f \to f \times \lsub$ and $d\sigma/dt \to d\sigma/dt /
\lsub$~\cite{Zhao:2020yvf}, under which the RBE is invariant. 
With the subdivision factor $\lsub>1$, this transformation decreases
the opacity by $\chi \to \chi /
\sqrt{\lsub}$~\cite{Zhang:1998tj,Zhao:2020yvf} and can thus reduce the 
causality violation. However, this parton subdivision method is
computationally expensive, because the parton number per event
increases as $N \to N \times \lsub$ and the number of collisions to
simulate each event increases by a factor of $\lsub^2$. 
Recently we found a new parton subdivision
transformation~\cite{Zhao:2020yvf} for particles in a box with
periodic boundary conditions, where we increase $f$ by decreasing the
volume as $V \to V / \lsub$. This new transformation is much faster
since the parton number per event remains the
same~\cite{Zhao:2020yvf}, but it can only be applied to box
simulations. Because the parton subdivision method changes the
event-by-event fluctuations and correlations~\cite{Zhao:2020yvf},
which are important observables for the study of nuclear collisions. 
we need a parton cascade that accurately solves the
RBE without using the parton subdivision method. 

In ZPC or any parton cascade, there are different scattering
prescriptions to implement the cascade method that generally lead to
different results~\cite{Kortemeyer:1995di,Zhang:1997ej}. 
A scattering prescription, or collision scheme, refers to the choice of
the collision frame, the collision space-time point(s), the ordering
frame, and the ordering time~\cite{Zhang:1997ej}. 
Whether a collision happens is determined in the collision frame
with the closest approach criterion, while possible collisions are
ordered according to the ordering time (in the ordering frame) and
performed in that order.  
The ZPC uses the two-parton center-of-momentum frame as the collision 
frame, while the ordering frame is taken as the rest frame of the
gas (i.e., the global frame)~\cite{Zhang:1997ej}. 
Although the collision time is a single value in the 
collision frame, the two partons generally have different spatial 
coordinates at that time due to the finite $\sigma$, which then lead
to two different times of collision $\tca$ and  $\tcb$ in the global
frame after Lorentz transformations~\cite{Zhao:2020yvf}. 
With the collision frame and ordering frame fixed, there are still 
different choices for the collision time and ordering time, leading to
different collision schemes. Note that each of the two partons
involved in a collision changes its  momentum at the chosen collision
time at its corresponding position in the global frame. 

Recently, it was shown that the inaccuracy from the causality
violation depends greatly on the collision scheme~\cite{Zhao:2020yvf}. 
In the original ZPC~\cite{Zhang:1997ej}, the ordering time and
collision times for both partons are all chosen as $(\tca+\tcb)/2$,
thus this original collision scheme is called the $t$-average scheme. 
Various collision schemes have been studied with ZPC, and a new
collision scheme was found to perform much better than the original 
scheme~\cite{Zhao:2020yvf}. 
The new scheme chooses the ordering time and the parton collision
times as ${\rm min}(\tca,\tcb)$, so it is called the $t$-minimum
scheme. This new collision scheme significantly improves the accuracy,
as judged from comparisons with the expected thermal distribution in
equilibrium as well as the time evolution of the momentum distribution
in comparison with the parton subdivision results~\cite{Zhao:2020yvf}. 
Since the latter depends on the assumption that the subdivision
results are correct, it is better to compare ZPC with an exact
analytical solution of the RBE.
While an analytical solution of the non-relativistic Boltzmann
equation for a homogeneous gas has long been
known~\cite{PhysRevLett.36.1107}, an exact analytical solution 
for a relativistic massless gas was found 
only recently~\cite{Bazow:2015dha,Bazow:2016oky}. 
Then the hadron transport SMASH was compared
with this exact solution, where good agreement was
observed~\cite{Tindall:2016try}. 
In this study, we assess and then improve the accuracy of ZPC
by comparing with this exact solution for the time evolution of the
full momentum distribution of a massless homogeneous gas. 

\section{Exact Analytical Solution of the RBE}

The recent exact analytical solution of the
RBE~\cite{Bazow:2015dha,Bazow:2016oky} for a massless, homogeneous and
isotropically expanding gas, which scatters with an energy-independent
and isotropic elastic cross section, is given by
\begin{equation}
	\begin{split}
f(p,\tau)&=\lambda\exp{\left[-\frac{u^\mu p_\mu}{T(\tau)\kappa(\tau)}\right]} \\
		&\times\left[\frac{4\kappa(\tau)-3}{\kappa^4(\tau)}
		+\frac{u^\mu p_\mu}{T(\tau)}\frac{1-\kappa(\tau)}{\kappa^5(\tau)}\right].
	\end{split}
	\label{ex-an-sol}
\end{equation}
Here, $p^\mu$ is the four-momentum, $\lambda$ is the
fugacity, $u^\mu$ is the four-velocity of the gas, and 
$\kappa(\tau)=1-\exp\left(-\tau/6\right)/4$.
The temperature $T(t)=T(0)/a(t)$ depends on a time-dependent scale
factor $a(t)$, which controls the spatial expansion in the 
Friedman-Lema\^itre-Robertson-Walker (FLRW) metric  
$ds^2=dt^2-a^2(t) (dx^2+dy^2+dz^2)$. 
The time variable is $\tau=\int_{\hat{t}_0}^{\hat{t}}dt'/a^3(t')$, 
where the scaled time is $\hat{t}=t/l$~\cite{Bazow:2015dha} with $l$
being the mean free path.  

The single-particle distribution function is independent of the
spatial coordinates and is spherically symmetric in the momentum
space: $f(x^\mu,p^\mu,\tau) \to f(p,\tau)$, where $p$ represents the
magnitude of momentum. The exact solution of Eq.~\eqref{ex-an-sol} has
the following initial condition corresponding to $\tau=0$ when
$T(0)=T_0$:
\begin{equation}
	f(p,0)=\lambda\frac{256}{243}\frac{p}{T_0}\exp\left(-\frac{4p}{3T_0}\right).
	\label{in-con}
\end{equation}

\section{Results}

The exact analytical solution is also valid for a constant scale
factor $a(t)=1$. 
So in this study we investigate this non-expanding case, which
corresponds to box calculations with periodic boundary 
conditions~\cite{Zhang:1997ej}.
Note that in this case the temperature of the system remains
constant at $T(\tau)=T$, and we choose $\lambda=1$ for the 
fugacity. Then the time evolution of Eq.~\eqref{ex-an-sol} can be
simplified as
\begin{equation}
		f(p,\tau) \!=\! \exp \! {\left[-\frac{p}{T\kappa(\tau)}\right]} 
		\left[\frac{4\kappa(\tau)-3}{\kappa^4(\tau)}
\!+\! \frac{p}{T} \frac{1-\kappa(\tau)}{\kappa^5(\tau)}\right].
\label{ex-sol}
\end{equation}
It approaches the thermal distribution at late times:
\begin{equation}
\feq(p) \equiv
f(p,\tau\to\infty)=\lambda\exp{\left(-\frac{p}{T}\right)}.
\label{f-eq}
\end{equation}

The evolution of a parton system depends on the density and
cross section and thus the system approaches equilibrium at different
rates~\cite{Zhao:2020yvf}. 
The scaled time $\hat{t}$~\cite{Bazow:2015dha} takes this into
account, where $\tau=\hat{t}$ for our choice of $a(t)=1$ and 
$\hat{t}_0=0$. 
For gluons under the Boltzmann statistics, the momentum distribution
is higher by the gluon degeneracy factor, and the opacity is given
by~\cite{Zhang:1998tj} 
\begin{equation}
	\chi = \frac{16}{\pi^2}T^3\sqrt{\frac{\sigma^3}{\pi}}.
\end{equation}
As a result, the time variable $\tau$ is related to the 
global time $t$ and opacity $\chi$ as
\begin{equation}
\tau=t~T\left (\frac{16\chi^2}{\pi} \right ) ^{\!\!\frac{1}{3}}.
\end{equation}
In this study, we run ZPC until an ending time of $\tau=60$ because
this time is long enough for each system to practically reach
equilibrium. For each configuration, we start ZPC at $\tau=0$ from the
initial condition of Eq.~\eqref{in-con}, and we use 1000 events with
16000 gluons per event to ensure that the results are statistically
reliable.

\begin{figure}
	\includegraphics[width=\linewidth]{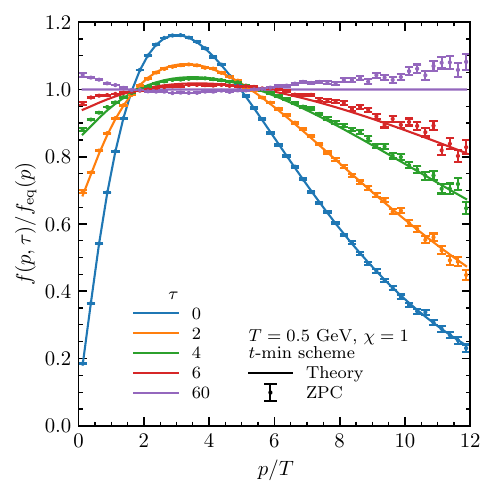}
	\caption{Ratio of $\fzpc(p,\tau)$ and $\fthe(p,\tau)$
	over the equilibrium distribution $\feq(p)$ versus $p/T$ at various 
        times for gluons in a box at $T=0.5$ GeV and $\sigma=1.65$ mb
        (for $\chi=1.0$), where $\fzpc(p,\tau)$ comes  from ZPC with the
        $t$-minimum collision scheme and $\fthe(p,\tau)$ represents
        the exact solution.
	\label{fig1}}
\end{figure}

In Fig.~\ref{fig1}, we examine the ZPC results from the $t$-minimum
scheme by plotting the ratio $f(p,\tau)/\feq(p)$ as a
function of $p/T$ at several $\tau$ during the evolution for a gluon
gas at $T=0.5$ GeV and $\sigma=1.65$ mb. 
The symbols represent the ratio of the ZPC distributions
$\fzpc(p,\tau)$ over the equilibrium distribution of Eq.~\eqref{f-eq},
while the solid lines represent the ratio of the analytical
distributions $\fthe(p,\tau)$ over the equilibrium distribution. Note
that the ZPC distributions are obtained at discrete values of $p/T$,
while the analytical (including the equilibrium) distributions are 
smooth functions of $p/T$. For apple-to-apple comparisons, 
for both $\fthe(p,\tau)$ and $\feq(p)$ we calculate the bin-averaged
value as $f(p_i)\equiv \int_a^b f(p)d^3p /\int_a^b d^3p$, where
$a=p_i-\Delta p/2$ and $b=p_i+\Delta p/2$ with a constant bin width
$\Delta p$.

We see in Fig.~\ref{fig1} that the initial distribution at $\tau=0$ is
higher than the equilibrium distribution for $p/T \in [1.64, 4.96]$
while lower in other $p/T$ ranges. 
Since we have implemented the initial condition of Eq.~\eqref{in-con} in
ZPC, $\fzpc(p,0)$ fully agrees with $\fthe(p,0)$ as intended. 
As $\tau$ increases, particle collisions cause the system 
to approach equilibrium, and the exact solution is seen to reach
equilibrium by $\tau=60$.  We see that the time evolution of the
ZPC momentum distributions at early time agrees very well with the exact
solution. At late times, however, the ZPC momentum distribution under
the $t$-minimum scheme shows a underpopulation within $p/T \in [1.6,
5.0]$ and an overpopulation elsewhere. 

\begin{figure*}
	\includegraphics[width=\linewidth]{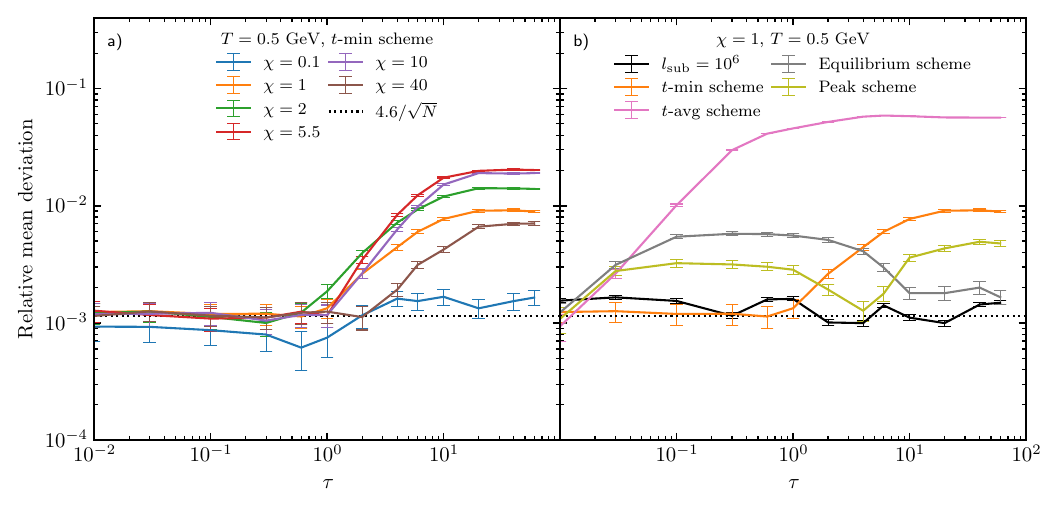}
	\caption{Time evolution of the relative mean deviation from ZPC
          for a gluon gas at $T=0.5$ GeV from (a) the
	$t$-minimum scheme at various $\chi$ values and 
        (b) several collision schemes at $\chi=1$.
	The subdivision result with $\lsub=10^6$ (black) and the
        expectation from pure statistical fluctuations (dotted) are
        also shown, and the values at $\tau=0$ have been 
        plotted at $\tau=10^{-2}$.
\label{fig2}}
\end{figure*}

To quantify the difference between $\fzpc(p,\tau)$ and $\fthe(p,\tau)$, we
calculate the relative mean deviation (RMD) between the two
distributions as 
\begin{equation}
	\MD(\tau) = 
	\sqrt{	\frac{\sum\limits_i\left[N_{\rm ZPC}(p_i,\tau)-N_{\rm
                Theory}(p_i,\tau)\right]^2} {\sum\limits_i\left[N_{\rm
                Theory}(p_i,\tau)\right]^2}}, 
	\label{stat}
\end{equation}
where $N(p_i,\tau) \propto f(p_i,\tau) \int_a^b d^3p
\appropto f(p_i,\tau) p_i^2$ is the number of partons in the $i_{\rm
  th}$ momentum bin from all events for the given configuration. 
For two distributions that differ at each momentum bin by the
same fraction ($\epsilon$ or -$\epsilon$), we would have
$\MD=\epsilon$.
On the other hand, if a momentum distribution (with $\Delta p=T/4$) is
different from the equilibrium distribution $\feq$ (or the initial
distribution)  only due to statistical fluctuations, we expect $\MD
\simeq 4.6/\sqrt{N}$ (or $4.4/\sqrt{N}$), where $N$ is the total
number of partons in all simulated events of a configuration (16 
million here). Note that these $\MD$ values due to statistical
fluctuations are $\propto 1/\sqrt{\Delta p/T}$, while the $\MD$ value
due to ``real'' differences in momentum distributions is
essentially independent of the bin width of $p/T$. 

In Fig.~\ref{fig2}(a), we compare the $\MD$ versus $\tau$
for ZPC calculations under the $t$-minimum scheme 
at several $\chi$ values (at $T=0.5$ GeV). 
At early times ($\tau < 1$), the ZPC results agree very well with the exact
solution results for all the shown opacities, and the deviations are 
consistent with the two differing by only statistical fluctuations. 
Figure~\ref{fig2}(a) also shows that the $\MD$
value generally increases with time for $\tau > 1$ until it reaches an
equilibrium value around $\tau=40$.  
Notably, the increase and equilibrium value of $\MD$ do not depend
monotonically on $\chi$, consistent with the behavior of $\vpT$ 
seen earlier~\cite{Zhao:2020yvf}.

\begin{figure}
	\includegraphics[width=\linewidth]{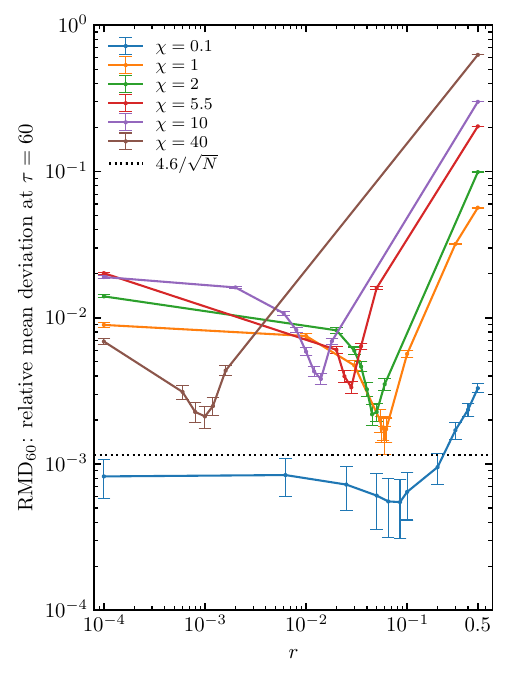}
	\caption{The relative mean deviation at equilibrium
          ($\tau=60$) versus the $r$ parameter for various
          opacities. Note that the values at $r=0$ are plotted at $r=10^{-4}$.
	\label{fig3}}
\end{figure}

To further improve the ZPC accuracy, we generalize the collision
scheme by introducing a parameter $r$ in 
the collision time (also chosen as the ordering time):
\begin{equation}
	t_c = {\rm min}(\tca,\tcb) + r \left\lvert \tca - \tcb \right \rvert.
	\label{col-sch}
\end{equation}
So the original ZPC $t$-average collision scheme corresponds to
$r=1/2$~\cite{Zhang:1997ej}, while the new $t$-minimum collision 
scheme corresponds to $r=0$~\cite{Zhao:2020yvf}. It is natural to
expect the collision time to be in-between $\tca$ and $\tcb$, i.e., $0 
\le r \le 1$.
We then run ZPC at a given $\chi$ value to find the $r$ value that
minimizes the relative mean deviation. Since the $\MD$ depends on the
evolution time, we choose to minimize either the equilibrium $\MD$ value at
$\tau=60$ ($\MD_{\rm eq}$) or the maximum $\MD$ value over all $\tau$
from 0 to 60 ($\MD_{\rm max}$);  we call the collision scheme
corresponding to the former as the equilibrium scheme and that
corresponding to the later as the peak scheme. 

We plot in Fig.~\ref{fig3} the $r$-dependence of $\MD$ at equilibrium
for various opacities. 
We see that for each opacity there is a $r$
value ($r_{\rm optimal}$) between 0 and 1/2 that minimizes the $\MD$,  
while the $r_{\rm optimal}$ value decreases as $\chi$ increases. 
In addition, the equilibrium $\MD$ value at $r=1/2$ is higher
than that at $r=0$ for each opacity, consistent with the earlier
finding that the $t$-minimum scheme is more accurate than
the original  $t$-average scheme~\cite{Zhao:2020yvf}. 
Note that the non-monotonic $\chi$-dependence of the equilibrium
$\MD$ for the $t$-minimum scheme is already shown in
Fig.~\ref{fig2}(a), and the non-zero $r_{\rm optimal}$ values shown in 
Fig.~\ref{fig3} mean that indeed the accuracy of ZPC can be further
improved. 
The $\MD_{\rm max}$ results versus $r$ for the peak collision scheme
are similar to Fig.~\ref{fig3}, except that the $r_{\rm optimal}$
values are smaller. 
Also note that the $\MD$ values at each opacity
shown here (and in Fig.~\ref{fig5}) correspond to the average of the
$\MD$ values for two temperatures $T=$ 0.2 and 0.5 GeV, although the 
results for the two temperatures are essentially the same.  
After obtaining the $r_{\rm optimal}$ values at
multiple opacities, we fit them and obtain the following
parametrization of $r_{\rm optimal}$ as a function of $\chi$ for both
the equilibrium scheme and peak scheme: 
\begin{equation}
r_{\rm optimal}(\chi) =
	\begin{cases}
		\frac{873}{(15.8+\chi)^{3.40}}, & \text{equilibrium scheme;} \\
		\frac{518}{(15.8+\chi)^{3.40}}, & \text{peak scheme.}
	\end{cases}
	\label{r-opt}
\end{equation}

In Fig.~\ref{fig2}(b), we show the $\tau$-evolution of the $\MD$ for
four collision schemes for a gluon gas at $T=0.5$ GeV and $\chi=1$.  
We also show the ZPC results using the new parton subdivision
method~\cite{Zhao:2020yvf} with the subdivision factor $\lsub=10^6$
as well as the expectation for pure statistical fluctuations. 
We see that both generalized collision schemes yield a significantly
lower $\MD$ value at equilibrium than the $t$-minimum scheme, while all
of them give $\MD$ values much lower than the original $t$-average scheme
at finite times (when $\tau>0.1$). We note that unfortunately the two
generalized collision schemes both give higher relative mean
deviations than the $t$-minimum scheme at intermediate times when
$\tau \in (0.03, 1.6)$.

\begin{figure}
	\includegraphics[width=\linewidth]{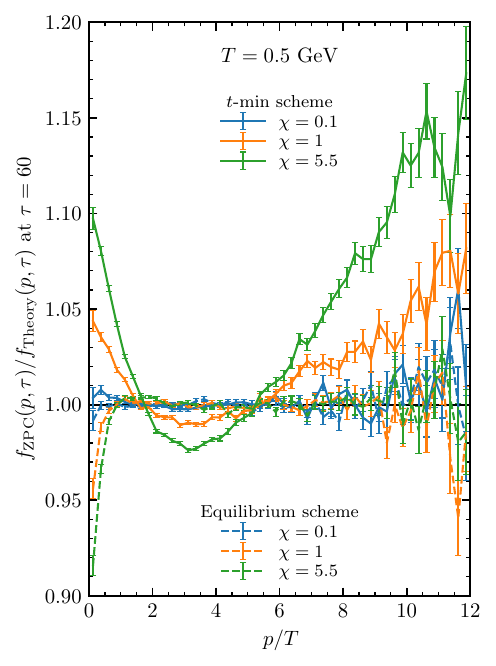}
	\caption{Ratio of the ZPC momentum distribution at equilibrium
          ($\tau=60$) over the theoretical distribution versus $p/T$
          at several opacities from the $t$-minimum scheme and equilibrium
          scheme for a gluon gas at $T=0.5$ GeV.
	\label{fig4}}
\end{figure}

Figure~\ref{fig4} shows the ratio $\fzpc(p,\tau)/\fthe(p,\tau)$
at equilibrium ($\tau=60$) for gluons in a box at $T=0.5$ GeV and
three opacities, where the ZPC results from both the $t$-minimum 
scheme and the equilibrium scheme of Eq.~\eqref{r-opt} are shown.  
Note that the theoretical $\fthe(p,\tau)$ values
have been bin-averaged as done for Fig.~\ref{fig1}. 
We see that the ZPC momentum distributions in equilibrium from 
both schemes are very close to the theoretical 
distribution at low opacity ($\chi=0.1$); this is expected since the
causality violation in the parton cascade at this opacity is very small. 
At high opacities, however, the ZPC distribution from the 
$t$-minimum scheme can deviate significantly from the theoretical 
distribution. 
We also see that the equilibrium scheme strongly suppresses the
deviations for most of the momentum range ($p/T>1$), although it tends
to underestimate the equilibrium distribution at $p/T \lesssim 1$.

\begin{figure}
	\includegraphics[width=\linewidth]{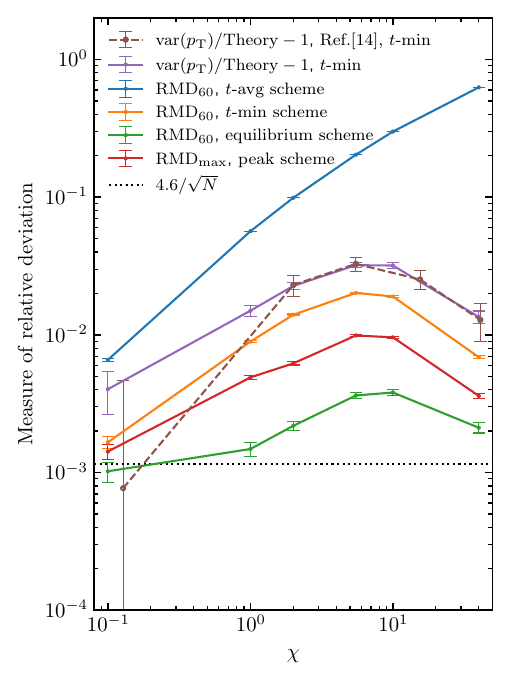}
	\caption{The relative mean deviation in equilibrium ($\MD_{60}$)
          for three collision schemes of ZPC, in addition to the
          relative deviation of the variance of the momentum distribution in
          equilibrium from the $t$-minimum scheme, as functions of the
          opacity. The maximum relative mean deviation for all $\tau$
          ($\MD_{\rm max}$) from the peak scheme (red) and the
          expectation from pure statistical fluctuations (dotted) are
          also shown.
	\label{fig5}}
\end{figure}

To summarize the accuracy of various collision schemes, 
we plot in Fig.~\ref{fig5} their $\MD$ values versus opacity. 
While the original $t$-average scheme shows a monotonic increase of 
the $\MD$ value at equilibrium from $\chi=0.1$ to $\chi=40$,  the
other three schemes all show a $\MD$ peak (around $\chi$ between 5 and
10) with much lower magnitudes. 
It is also clear that the two generalized schemes of Eq.~\eqref{r-opt}
are more accurate than the $t$-minimum scheme. For moderate to high 
opacities ($\chi \geq 1$), the equilibrium scheme reduces the
equilibrium $\MD$ value by a factor of $3$ to $6$. 
The red curve shows the maximum $\MD$ value (throughout the time
evolution) from the peak scheme for each opacity. 
Since the maximum $\MD$ value for the $t$-minimum scheme always
happens at or close to equilibrium (as shown in Fig.~\ref{fig2}), 
Fig.~\ref{fig5} shows that the peak scheme reduces 
the maximum $\MD$ value by a factor of $\sim 2$
at moderate to high opacities ($\chi \geq 1$) when compared to the
$t$-minimum scheme.

Figure~\ref{fig5} also shows the relative difference between the
variance of the ZPC $\pT$ distribution in equilibrium from the
$t$-minimum scheme and the expectation from the thermal distribution.  
Here, the variance is $\vpT=\langle\pT^2\rangle-\mpT^2$, and its
theoretical expectation is $(8-9\pi^2/16)T^2$ for massless
partons under the Boltzmann distribution. 
For comparison, the $\vpT$ results from an earlier ZPC
study~\cite{Zhao:2020yvf} are also shown here, which
agree with our current results after considering the
estimated error bars of the earlier results. 
We see that the relative difference of $\vpT$ versus opacity 
from the $t$-minimum scheme closely follows the shape of the relative
mean deviation but are higher in magnitude. 
Like the relative mean deviations for three out of four collision schemes, 
the relative difference of $\vpT$ also peaks at $\chi$ between 5
and 10. 

\section{Conclusion}
 
Transport models based on the kinetic theory are important tools for the 
study of non-equilibrium dynamics in nuclear collisions including the
origin of collectivity from small to large colliding systems. It is
thus essential for a parton transport to be able to solve the
relativistic Boltzmann equation accurately. 
Preferably this should be achieved without using the parton subdivision
method, which alters the event-by-event correlations and
fluctuations. In this study, we first examine and then improve 
the ZPC parton cascade for gluons under isotropic elastic
scatterings in a box with periodic boundary conditions. 
By comparing the results with an exact analytical solution of
the relativistic Boltzmann equation for a given initial condition, we
can for the first time quantify the parton cascade accuracy 
in the full momentum distribution throughout the time evolution. 
We first confirm that, compared to the original
$t$-average collision scheme, the recent $t$-minimum collision scheme
significantly improves the accuracy, leading to a relative mean deviation of
$\lesssim \! 2\%$ in equilibrium for all opacities. 
We then generalize the collision scheme to be opacity-dependent
to further improve the accuracy, where the peak scheme has 
a relative mean deviation of $\leq \! 1\%$ 
for the momentum distribution at any time for all opacities (up to
$\chi=40$). These levels of accuracy at high opacities ($\chi>1$) are
noteworthy because naively the parton cascade is expected to fail
there.  An accurate parton cascade for box calculations lays the
foundation to apply and assess the parton cascade in more realistic 
3-dimensional expansion cases. 

\section*{Acknowledgement}
This work has been supported by the National Science Foundation under
Grant No. 2012947 and 2310021.
We thank Dr. M. Martinez for discussions about the exact solution. We
would like to acknowledge the use of the following software:
Matplotlib~\cite{Hunter:2007}, Numpy~\cite{Harris:2020xlr}, Scipy~\cite{Virtanen:2019joe}. 

\bibliography{exact-solution}

\begin{thebibliography}{37}
\expandafter\ifx\csname natexlab\endcsname\relax\def\natexlab#1{#1}\fi
\expandafter\ifx\csname bibnamefont\endcsname\relax
  \def\bibnamefont#1{#1}\fi
\expandafter\ifx\csname bibfnamefont\endcsname\relax
  \def\bibfnamefont#1{#1}\fi
\expandafter\ifx\csname citenamefont\endcsname\relax
  \def\citenamefont#1{#1}\fi
\expandafter\ifx\csname url\endcsname\relax
  \def\url#1{\texttt{#1}}\fi
\expandafter\ifx\csname urlprefix\endcsname\relax\def\urlprefix{URL }\fi
\providecommand{\bibinfo}[2]{#2}
\providecommand{\eprint}[2][]{\url{#2}}

\bibitem[{\citenamefont{Adams et~al.}(2005)}]{STAR:2005gfr}
\bibinfo{author}{\bibfnamefont{J.}~\bibnamefont{Adams}} \bibnamefont{et~al.}
  (\bibinfo{collaboration}{STAR}), \bibinfo{journal}{Nucl. Phys. A}
  \textbf{\bibinfo{volume}{757}}, \bibinfo{pages}{102} (\bibinfo{year}{2005}).

\bibitem[{\citenamefont{Abelev et~al.}(2014)}]{ALICE:2013nwm}
\bibinfo{author}{\bibfnamefont{B.}~\bibnamefont{Abelev}} \bibnamefont{et~al.}
  (\bibinfo{collaboration}{ALICE}), \bibinfo{journal}{J. Phys. G}
  \textbf{\bibinfo{volume}{41}}, \bibinfo{pages}{087002}
  (\bibinfo{year}{2014}).

\bibitem[{\citenamefont{Abdallah et~al.}(2022)}]{STAR:2021yiu}
\bibinfo{author}{\bibfnamefont{M.~S.} \bibnamefont{Abdallah}}
  \bibnamefont{et~al.} (\bibinfo{collaboration}{STAR}), \bibinfo{journal}{Phys.
  Lett. B} \textbf{\bibinfo{volume}{827}}, \bibinfo{pages}{137003}
  (\bibinfo{year}{2022}).

\bibitem[{\citenamefont{STAR}(2025)}]{STAR:2025owm}
\bibinfo{author}{\bibnamefont{STAR}} (\bibinfo{year}{2025}),
  \eprint{2504.02531}.

\bibitem[{\citenamefont{Molnar and Voloshin}(2003)}]{Molnar:2003ff}
\bibinfo{author}{\bibfnamefont{D.}~\bibnamefont{Molnar}} \bibnamefont{and}
  \bibinfo{author}{\bibfnamefont{S.~A.} \bibnamefont{Voloshin}},
  \bibinfo{journal}{Phys. Rev. Lett.} \textbf{\bibinfo{volume}{91}},
  \bibinfo{pages}{092301} (\bibinfo{year}{2003}).

\bibitem[{\citenamefont{Adare et~al.}(2015)}]{PHENIX:2014fnc}
\bibinfo{author}{\bibfnamefont{A.}~\bibnamefont{Adare}} \bibnamefont{et~al.}
  (\bibinfo{collaboration}{PHENIX}), \bibinfo{journal}{Phys. Rev. Lett.}
  \textbf{\bibinfo{volume}{114}}, \bibinfo{pages}{192301}
  (\bibinfo{year}{2015}).

\bibitem[{\citenamefont{Aidala et~al.}(2019)}]{PHENIX:2018lia}
\bibinfo{author}{\bibfnamefont{C.}~\bibnamefont{Aidala}} \bibnamefont{et~al.}
  (\bibinfo{collaboration}{PHENIX}), \bibinfo{journal}{Nature Phys.}
  \textbf{\bibinfo{volume}{15}}, \bibinfo{pages}{214} (\bibinfo{year}{2019}).

\bibitem[{\citenamefont{Abdulhamid et~al.}(2023)}]{STAR:2022pfn}
\bibinfo{author}{\bibfnamefont{M.~I.} \bibnamefont{Abdulhamid}}
  \bibnamefont{et~al.} (\bibinfo{collaboration}{STAR}), \bibinfo{journal}{Phys.
  Rev. Lett.} \textbf{\bibinfo{volume}{130}}, \bibinfo{pages}{242301}
  (\bibinfo{year}{2023}).

\bibitem[{\citenamefont{Khachatryan et~al.}(2010)}]{CMS:2010ifv}
\bibinfo{author}{\bibfnamefont{V.}~\bibnamefont{Khachatryan}}
  \bibnamefont{et~al.} (\bibinfo{collaboration}{CMS}), \bibinfo{journal}{JHEP}
  \textbf{\bibinfo{volume}{09}}, \bibinfo{pages}{091} (\bibinfo{year}{2010}).

\bibitem[{\citenamefont{Khachatryan et~al.}(2015)}]{CMS:2015yux}
\bibinfo{author}{\bibfnamefont{V.}~\bibnamefont{Khachatryan}}
  \bibnamefont{et~al.} (\bibinfo{collaboration}{CMS}), \bibinfo{journal}{Phys.
  Rev. Lett.} \textbf{\bibinfo{volume}{115}}, \bibinfo{pages}{012301}
  (\bibinfo{year}{2015}).

\bibitem[{\citenamefont{Heller and Spalinski}(2015)}]{Heller:2015dha}
\bibinfo{author}{\bibfnamefont{M.~P.} \bibnamefont{Heller}} \bibnamefont{and}
  \bibinfo{author}{\bibfnamefont{M.}~\bibnamefont{Spalinski}},
  \bibinfo{journal}{Phys. Rev. Lett.} \textbf{\bibinfo{volume}{115}},
  \bibinfo{pages}{072501} (\bibinfo{year}{2015}).

\bibitem[{\citenamefont{Kurkela et~al.}(2018)\citenamefont{Kurkela, Wiedemann,
  and Wu}}]{Kurkela:2018ygx}
\bibinfo{author}{\bibfnamefont{A.}~\bibnamefont{Kurkela}},
  \bibinfo{author}{\bibfnamefont{U.~A.} \bibnamefont{Wiedemann}},
  \bibnamefont{and} \bibinfo{author}{\bibfnamefont{B.}~\bibnamefont{Wu}},
  \bibinfo{journal}{Phys. Lett. B} \textbf{\bibinfo{volume}{783}},
  \bibinfo{pages}{274} (\bibinfo{year}{2018}).

\bibitem[{\citenamefont{Kurkela et~al.}(2020)\citenamefont{Kurkela, van~der
  Schee, Wiedemann, and Wu}}]{Kurkela:2019set}
\bibinfo{author}{\bibfnamefont{A.}~\bibnamefont{Kurkela}},
  \bibinfo{author}{\bibfnamefont{W.}~\bibnamefont{van~der Schee}},
  \bibinfo{author}{\bibfnamefont{U.~A.} \bibnamefont{Wiedemann}},
  \bibnamefont{and} \bibinfo{author}{\bibfnamefont{B.}~\bibnamefont{Wu}},
  \bibinfo{journal}{Phys. Rev. Lett.} \textbf{\bibinfo{volume}{124}},
  \bibinfo{pages}{102301} (\bibinfo{year}{2020}).

\bibitem[{\citenamefont{He et~al.}(2016)\citenamefont{He, Edmonds, Lin, Liu,
  Molnar, and Wang}}]{He:2015hfa}
\bibinfo{author}{\bibfnamefont{L.}~\bibnamefont{He}},
  \bibinfo{author}{\bibfnamefont{T.}~\bibnamefont{Edmonds}},
  \bibinfo{author}{\bibfnamefont{Z.-W.} \bibnamefont{Lin}},
  \bibinfo{author}{\bibfnamefont{F.}~\bibnamefont{Liu}},
  \bibinfo{author}{\bibfnamefont{D.}~\bibnamefont{Molnar}}, \bibnamefont{and}
  \bibinfo{author}{\bibfnamefont{F.}~\bibnamefont{Wang}},
  \bibinfo{journal}{Phys. Lett. B} \textbf{\bibinfo{volume}{753}},
  \bibinfo{pages}{506} (\bibinfo{year}{2016}).

\bibitem[{\citenamefont{Lin et~al.}(2016)\citenamefont{Lin, He, Edmonds, Liu,
  Molnar, and Wang}}]{Lin:2015ucn}
\bibinfo{author}{\bibfnamefont{Z.-W.} \bibnamefont{Lin}},
  \bibinfo{author}{\bibfnamefont{L.}~\bibnamefont{He}},
  \bibinfo{author}{\bibfnamefont{T.}~\bibnamefont{Edmonds}},
  \bibinfo{author}{\bibfnamefont{F.}~\bibnamefont{Liu}},
  \bibinfo{author}{\bibfnamefont{D.}~\bibnamefont{Molnar}}, \bibnamefont{and}
  \bibinfo{author}{\bibfnamefont{F.}~\bibnamefont{Wang}},
  \bibinfo{journal}{Nucl. Phys. A} \textbf{\bibinfo{volume}{956}},
  \bibinfo{pages}{316} (\bibinfo{year}{2016}).

\bibitem[{\citenamefont{Li et~al.}(2024)\citenamefont{Li, Song, Xu, Sun, and
  Wang}}]{Li:2024ivj}
\bibinfo{author}{\bibfnamefont{K.}~\bibnamefont{Li}},
  \bibinfo{author}{\bibfnamefont{H.-F.} \bibnamefont{Song}},
  \bibinfo{author}{\bibfnamefont{H.-J.} \bibnamefont{Xu}},
  \bibinfo{author}{\bibfnamefont{Y.-L.} \bibnamefont{Sun}}, \bibnamefont{and}
  \bibinfo{author}{\bibfnamefont{F.}~\bibnamefont{Wang}}
  (\bibinfo{year}{2024}), \eprint{2405.02847}.

\bibitem[{\citenamefont{Zhang}(1998)}]{Zhang:1997ej}
\bibinfo{author}{\bibfnamefont{B.}~\bibnamefont{Zhang}},
  \bibinfo{journal}{Comput. Phys. Commun.} \textbf{\bibinfo{volume}{109}},
  \bibinfo{pages}{193} (\bibinfo{year}{1998}).

\bibitem[{\citenamefont{Xu and Greiner}(2005)}]{Xu:2004mz}
\bibinfo{author}{\bibfnamefont{Z.}~\bibnamefont{Xu}} \bibnamefont{and}
  \bibinfo{author}{\bibfnamefont{C.}~\bibnamefont{Greiner}},
  \bibinfo{journal}{Phys. Rev. C} \textbf{\bibinfo{volume}{71}},
  \bibinfo{pages}{064901} (\bibinfo{year}{2005}).

\bibitem[{\citenamefont{Kurkela et~al.}(2024)\citenamefont{Kurkela,
  T{\"o}rnkvist, and Zapp}}]{Kurkela:2022qhn}
\bibinfo{author}{\bibfnamefont{A.}~\bibnamefont{Kurkela}},
  \bibinfo{author}{\bibfnamefont{R.}~\bibnamefont{T{\"o}rnkvist}},
  \bibnamefont{and} \bibinfo{author}{\bibfnamefont{K.}~\bibnamefont{Zapp}},
  \bibinfo{journal}{Eur. Phys. J. C} \textbf{\bibinfo{volume}{84}},
  \bibinfo{pages}{74} (\bibinfo{year}{2024}).

\bibitem[{\citenamefont{Kurkela et~al.}(2019)\citenamefont{Kurkela,
  Mazeliauskas, Paquet, Schlichting, and Teaney}}]{Kurkela:2018vqr}
\bibinfo{author}{\bibfnamefont{A.}~\bibnamefont{Kurkela}},
  \bibinfo{author}{\bibfnamefont{A.}~\bibnamefont{Mazeliauskas}},
  \bibinfo{author}{\bibfnamefont{J.-F.} \bibnamefont{Paquet}},
  \bibinfo{author}{\bibfnamefont{S.}~\bibnamefont{Schlichting}},
  \bibnamefont{and} \bibinfo{author}{\bibfnamefont{D.}~\bibnamefont{Teaney}},
  \bibinfo{journal}{Phys. Rev. C} \textbf{\bibinfo{volume}{99}},
  \bibinfo{pages}{034910} (\bibinfo{year}{2019}).

\bibitem[{\citenamefont{Sorge et~al.}(1989)\citenamefont{Sorge, Stoecker, and
  Greiner}}]{Sorge:1989dy}
\bibinfo{author}{\bibfnamefont{H.}~\bibnamefont{Sorge}},
  \bibinfo{author}{\bibfnamefont{H.}~\bibnamefont{Stoecker}}, \bibnamefont{and}
  \bibinfo{author}{\bibfnamefont{W.}~\bibnamefont{Greiner}},
  \bibinfo{journal}{Annals Phys.} \textbf{\bibinfo{volume}{192}},
  \bibinfo{pages}{266} (\bibinfo{year}{1989}).

\bibitem[{\citenamefont{Bass et~al.}(1998)}]{Bass:1998ca}
\bibinfo{author}{\bibfnamefont{S.~A.} \bibnamefont{Bass}} \bibnamefont{et~al.},
  \bibinfo{journal}{Prog. Part. Nucl. Phys.} \textbf{\bibinfo{volume}{41}},
  \bibinfo{pages}{255} (\bibinfo{year}{1998}).

\bibitem[{\citenamefont{Janka et~al.}(2007)\citenamefont{Janka, Langanke,
  Marek, Martinez-Pinedo, and Mueller}}]{Janka:2006fh}
\bibinfo{author}{\bibfnamefont{H.-T.} \bibnamefont{Janka}},
  \bibinfo{author}{\bibfnamefont{K.}~\bibnamefont{Langanke}},
  \bibinfo{author}{\bibfnamefont{A.}~\bibnamefont{Marek}},
  \bibinfo{author}{\bibfnamefont{G.}~\bibnamefont{Martinez-Pinedo}},
  \bibnamefont{and} \bibinfo{author}{\bibfnamefont{B.}~\bibnamefont{Mueller}},
  \bibinfo{journal}{Phys. Rept.} \textbf{\bibinfo{volume}{442}},
  \bibinfo{pages}{38} (\bibinfo{year}{2007}).

\bibitem[{\citenamefont{Florkowski et~al.}(2013)\citenamefont{Florkowski,
  Ryblewski, and Strickland}}]{Florkowski:2013lza}
\bibinfo{author}{\bibfnamefont{W.}~\bibnamefont{Florkowski}},
  \bibinfo{author}{\bibfnamefont{R.}~\bibnamefont{Ryblewski}},
  \bibnamefont{and}
  \bibinfo{author}{\bibfnamefont{M.}~\bibnamefont{Strickland}},
  \bibinfo{journal}{Nucl. Phys. A} \textbf{\bibinfo{volume}{916}},
  \bibinfo{pages}{249} (\bibinfo{year}{2013}).

\bibitem[{\citenamefont{Kodama et~al.}(1984)\citenamefont{Kodama, Duarte,
  Chung, Donangelo, and Nazareth}}]{Kodama:1983yk}
\bibinfo{author}{\bibfnamefont{T.}~\bibnamefont{Kodama}},
  \bibinfo{author}{\bibfnamefont{S.~B.} \bibnamefont{Duarte}},
  \bibinfo{author}{\bibfnamefont{K.~C.} \bibnamefont{Chung}},
  \bibinfo{author}{\bibfnamefont{R.}~\bibnamefont{Donangelo}},
  \bibnamefont{and} \bibinfo{author}{\bibfnamefont{R.~A. M.~S.}
  \bibnamefont{Nazareth}}, \bibinfo{journal}{Phys. Rev. C}
  \textbf{\bibinfo{volume}{29}}, \bibinfo{pages}{2146} (\bibinfo{year}{1984}).

\bibitem[{\citenamefont{Kortemeyer et~al.}(1995)\citenamefont{Kortemeyer,
  Bauer, Haglin, Murray, and Pratt}}]{Kortemeyer:1995di}
\bibinfo{author}{\bibfnamefont{G.}~\bibnamefont{Kortemeyer}},
  \bibinfo{author}{\bibfnamefont{W.}~\bibnamefont{Bauer}},
  \bibinfo{author}{\bibfnamefont{K.}~\bibnamefont{Haglin}},
  \bibinfo{author}{\bibfnamefont{J.}~\bibnamefont{Murray}}, \bibnamefont{and}
  \bibinfo{author}{\bibfnamefont{S.}~\bibnamefont{Pratt}},
  \bibinfo{journal}{Phys. Rev. C} \textbf{\bibinfo{volume}{52}},
  \bibinfo{pages}{2714} (\bibinfo{year}{1995}).

\bibitem[{\citenamefont{T{\"o}rnkvist and Zapp}(2024)}]{Tornkvist:2023kan}
\bibinfo{author}{\bibfnamefont{R.}~\bibnamefont{T{\"o}rnkvist}}
  \bibnamefont{and} \bibinfo{author}{\bibfnamefont{K.}~\bibnamefont{Zapp}},
  \bibinfo{journal}{Phys. Lett. B} \textbf{\bibinfo{volume}{856}},
  \bibinfo{pages}{138952} (\bibinfo{year}{2024}).

\bibitem[{\citenamefont{Zhang et~al.}(1998)\citenamefont{Zhang, Gyulassy, and
  Pang}}]{Zhang:1998tj}
\bibinfo{author}{\bibfnamefont{B.}~\bibnamefont{Zhang}},
  \bibinfo{author}{\bibfnamefont{M.}~\bibnamefont{Gyulassy}}, \bibnamefont{and}
  \bibinfo{author}{\bibfnamefont{Y.}~\bibnamefont{Pang}},
  \bibinfo{journal}{Phys. Rev. C} \textbf{\bibinfo{volume}{58}},
  \bibinfo{pages}{1175} (\bibinfo{year}{1998}).

\bibitem[{\citenamefont{Molnar and Gyulassy}(2000)}]{Molnar:2000jh}
\bibinfo{author}{\bibfnamefont{D.}~\bibnamefont{Molnar}} \bibnamefont{and}
  \bibinfo{author}{\bibfnamefont{M.}~\bibnamefont{Gyulassy}},
  \bibinfo{journal}{Phys. Rev. C} \textbf{\bibinfo{volume}{62}},
  \bibinfo{pages}{054907} (\bibinfo{year}{2000}).

\bibitem[{\citenamefont{Zhao et~al.}(2020)\citenamefont{Zhao, Ma, Ma, and
  Lin}}]{Zhao:2020yvf}
\bibinfo{author}{\bibfnamefont{X.-L.} \bibnamefont{Zhao}},
  \bibinfo{author}{\bibfnamefont{G.-L.} \bibnamefont{Ma}},
  \bibinfo{author}{\bibfnamefont{Y.-G.} \bibnamefont{Ma}}, \bibnamefont{and}
  \bibinfo{author}{\bibfnamefont{Z.-W.} \bibnamefont{Lin}},
  \bibinfo{journal}{Phys. Rev. C} \textbf{\bibinfo{volume}{102}},
  \bibinfo{pages}{024904} (\bibinfo{year}{2020}).

\bibitem[{\citenamefont{Krook and Wu}(1976)}]{PhysRevLett.36.1107}
\bibinfo{author}{\bibfnamefont{M.}~\bibnamefont{Krook}} \bibnamefont{and}
  \bibinfo{author}{\bibfnamefont{T.~T.} \bibnamefont{Wu}},
  \bibinfo{journal}{Phys. Rev. Lett.} \textbf{\bibinfo{volume}{36}},
  \bibinfo{pages}{1107} (\bibinfo{year}{1976}).

\bibitem[{\citenamefont{Bazow et~al.}(2016{\natexlab{a}})\citenamefont{Bazow,
  Denicol, Heinz, Martinez, and Noronha}}]{Bazow:2015dha}
\bibinfo{author}{\bibfnamefont{D.}~\bibnamefont{Bazow}},
  \bibinfo{author}{\bibfnamefont{G.~S.} \bibnamefont{Denicol}},
  \bibinfo{author}{\bibfnamefont{U.}~\bibnamefont{Heinz}},
  \bibinfo{author}{\bibfnamefont{M.}~\bibnamefont{Martinez}}, \bibnamefont{and}
  \bibinfo{author}{\bibfnamefont{J.}~\bibnamefont{Noronha}},
  \bibinfo{journal}{Phys. Rev. Lett.} \textbf{\bibinfo{volume}{116}},
  \bibinfo{pages}{022301} (\bibinfo{year}{2016}{\natexlab{a}}).

\bibitem[{\citenamefont{Bazow et~al.}(2016{\natexlab{b}})\citenamefont{Bazow,
  Denicol, Heinz, Martinez, and Noronha}}]{Bazow:2016oky}
\bibinfo{author}{\bibfnamefont{D.}~\bibnamefont{Bazow}},
  \bibinfo{author}{\bibfnamefont{G.~S.} \bibnamefont{Denicol}},
  \bibinfo{author}{\bibfnamefont{U.}~\bibnamefont{Heinz}},
  \bibinfo{author}{\bibfnamefont{M.}~\bibnamefont{Martinez}}, \bibnamefont{and}
  \bibinfo{author}{\bibfnamefont{J.}~\bibnamefont{Noronha}},
  \bibinfo{journal}{Phys. Rev. D} \textbf{\bibinfo{volume}{94}},
  \bibinfo{pages}{125006} (\bibinfo{year}{2016}{\natexlab{b}}).

\bibitem[{\citenamefont{Tindall et~al.}(2017)\citenamefont{Tindall,
  Torres-Rincon, Rose, and Petersen}}]{Tindall:2016try}
\bibinfo{author}{\bibfnamefont{J.}~\bibnamefont{Tindall}},
  \bibinfo{author}{\bibfnamefont{J.~M.} \bibnamefont{Torres-Rincon}},
  \bibinfo{author}{\bibfnamefont{J.~B.} \bibnamefont{Rose}}, \bibnamefont{and}
  \bibinfo{author}{\bibfnamefont{H.}~\bibnamefont{Petersen}},
  \bibinfo{journal}{Phys. Lett. B} \textbf{\bibinfo{volume}{770}},
  \bibinfo{pages}{532} (\bibinfo{year}{2017}).

\bibitem[{\citenamefont{Hunter}(2007)}]{Hunter:2007}
\bibinfo{author}{\bibfnamefont{J.~D.} \bibnamefont{Hunter}},
  \bibinfo{journal}{Computing in Science \& Engineering}
  \textbf{\bibinfo{volume}{9}}, \bibinfo{pages}{90} (\bibinfo{year}{2007}).

\bibitem[{\citenamefont{Harris et~al.}(2020)}]{Harris:2020xlr}
\bibinfo{author}{\bibfnamefont{C.~R.} \bibnamefont{Harris}}
  \bibnamefont{et~al.}, \bibinfo{journal}{Nature}
  \textbf{\bibinfo{volume}{585}}, \bibinfo{pages}{357} (\bibinfo{year}{2020}).

\bibitem[{\citenamefont{Virtanen et~al.}(2020)}]{Virtanen:2019joe}
\bibinfo{author}{\bibfnamefont{P.}~\bibnamefont{Virtanen}}
  \bibnamefont{et~al.}, \bibinfo{journal}{Nature Meth.}
  \textbf{\bibinfo{volume}{17}}, \bibinfo{pages}{261} (\bibinfo{year}{2020}).

\end{thebibliography}

\end{document}